# Radioactive decay simulation with Geant4: experimental benchmarks and developments for X-ray astronomy applications

Steffen HAUF[1*], Markus KUSTER[1], Dieter H. H. HOFFMANN[1], Zane W. BELL[2], Maria Grazia PIA[3], Georg WEIDENSPOINTNER[4], Andreas ZOGLAUER[5]

[1]*TU Darmstadt, Darmstadt, Germany*
[2]*ORNL, Oak Ridge, TN 37830, USA*
[3]*INFN Sezione di Genova, 16146 Genova, Italy*
[4]*MPE and MPI Halbleiterlabor, 81739 München, Germany*
[5]*University of California at Berkeley, 94720 Berkeley, CA, USA*

We present γ spectroscopy validation measurements for the Geant4 radioactive decay simulation for a selected range of isotopes using a simple experimental setup. Using these results we point out problems in the decay simulation and where they may originate from.

## I. Introduction

Most current HEP (High Energy Physics) and astrophysical experiments today rely on extensive simulation of instrument response and background in their planning and operation phases. When high energy particles are of interest, this is often done using Monte Carlo simulation tools such as Geant4[1,2] or Fluka[3,4], which have been tested on a variety of experiments. As the energy and sensitivity of experiments increase, an accurate estimate of radiation dose and background due to radioactive decays in or near the detectors becomes more and more important.

In the case of the Geant4 Monte Carlo tool kit radioactive decay simulation can be included as a separate physical process which is part of the main distribution. A variety of works[5,6,7,8] have used the simulation to determine the efficiency of various detectors. Naturally the authors of these publications have focused on the prominent peaks within the respective spectra for efficiency estimates, which were often found in good agreement with experimental data. Because of this, thorough validations of the γ continuum are scarce. This is a problem for experiments such the Wide Field Imager[9] aboard the International X-ray Observatory, which rely on detailed knowledge of the complete decay spectrum to estimate the instrument response and background over their entire energy range.

We have conducted gamma spectroscopy measurements using $^{22}$Na, $^{133}$Na, $^{57}$Co, $^{60}$Co, $^{54}$Mn, $^{137}$Cs, which possess a variety of continuum and peak features between 20 and 1500 keV. From the comparison of these measurements with a Geant4 simulation of the experiment, problems within the simulation are identified and discussed.

---

*Corresponding Author, hauf@astropp.physik.tu-darmstadt.de.

## II. Experimental Overview

To minimize the effect of experimental uncertainties the experimental setup was kept simple. As shown in **Figure 1**, the experiment consisted of an ORTEC HPGe detector with an end cap size of 70 mm and a 70 nm thick Beryllium entrance window, which was shielded by lead and way blocks. Directly around the detector additional shielding in form of aluminum and copper tubes were installed. This setup was placed inside hollowed lead blocks.

The radioactive sources used were "Laboratory Type-D" discs whose activity was known to be 37,000 Bq in June 2006. These were suspended by a wire in front of the detector. The sources were measured for 10 minutes, a 1 hour background measurement was done before and after the measurements.

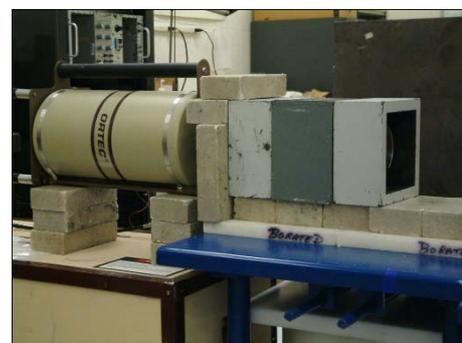

**Fig. 1**  The experimental setup of the HPGe detector. The detector is enclosed in the lead blocks shown on the right. The source was suspended from a string directly in front of it. A small part of the aluminum and copper tubes is also visible inside the lead box.

## III. Simulations Overview

For the simulations the experimental environment was simplified where applicable. This especially includes the wax and lead shielding. The sensitive detector was modeled as bulk Germanium with the depleted layers omitted. A dead layer of variable thickness which was not sensitive was included between the Beryllium window and the detector in order to test its effects on the overall detector response. The inner copper and aluminum shielding was modeled as simple hollow tube, open on both ends. The lead blocks consist of CSG (Constructed Solid Geometry) cubes.

The radioactive source was modeled according to the dimensions shown in **Figure 2** which were taken from the manufacturers technical description. The active part consisted of the decaying element which was enclosed in a PE disc. The general particle source (GPS) was set up to emit the decaying isotopes inside the radioactive volume. In this way self absorption within the disc was taken into account.

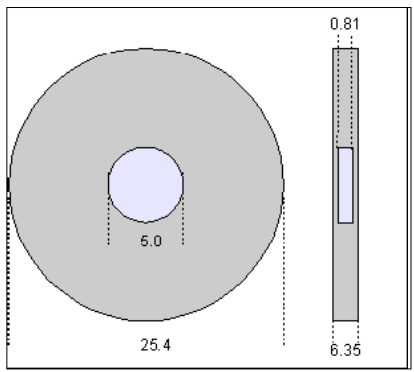

**Fig. 2** Schematic of the source disc. All dimensions are given in mm and are not to scale. The inner volume consists of the radioactive isotope, the outer part is plastic.

The simulation output consisted of the total energy per event deposited in the sensitive HPGe volume, the energy deposited by electrons and gammas as well as the total energy spectrum and the constituents energy spectrum before entering the beryllium entrance window. A more detailed output also included information about all the processes and particles involved and their geometrical origins. While the simple output was directed into ASCII text files, the detailed output was saved in binary format.

The stability of the simulation environment was tested using Geant4 versions 4.9.1, 4.9.1 with Doppler broadening and 4.9.2 with or without Low Energy EM extensions respectively. The default production cuts were generally set to 0.01 mm in order to accurately simulate the thin beryllium entrance window. The impact of the cut setting and the prodfac parameter of multiple scatting on the simulation was tested by varying these parameters. Hadronic physics could be turned on and off. These simulations usually involved simulating 10 million decays of the source isotope. From the outcome of theses simulations Geant4 4.9.1 with Doppler broadening was chosen as the basis for more elaborate simulation runs. This was done primarily due to the fact that the Low Energy extensions for this version have been validated with experimental data. These longer simulations consisted of at least 20 million decays up to 100 million, if an accurate representation of the continuum at very low or very high energies was need. Since our stability simulations showed that the influence of the hadronic processes is minimal, these longer simulations were done without hadronic physics, which greatly improves simulation speed.

## V. Analysis

The data analysis routines were programmed in the IDL programming language. They include spectral comparison of the experimental and simulated measurements along with Anderson-Darling and Kolmogorov goodness of fit tests as well as visualization of the processes and particles involved. The possibility of outputting the geometrical origin of the simulated particles was also included.

For analysis the total experimental and simulated spectra were generally binned into 0.5 keV energy intervals near the native experimental binning of around 0.3 keV, linearly increasing with energy when focusing on the peak features. For goodness of fits test the bin size was increased to 2 keV bins in order to reduce the impact of statistical fluctuations.

The simulated spectra were normalized to the experimental data by using the known activity of the experimentally measured isotopes. From this the normalization factor can be easily obtained from

$$\quad$$

with $N_{nominal}$ being the known activity at a specified date, $\lambda$ being the decay parameter, $t$ the time passed when the experiment was done and $N_{simulated}$ the number of decays simulated. Since both the experimental source and the simulated source isotropically decay at an opening angle of $4\pi$ and the detectors have the same distance and window size, the geometry of the experiment does not have to be accounted for. The background was subtracted for the experimental spectra, since the Geant4 simulations do not include this background.

## V. Results

### 1. Simulation stability, Influence of physics processes

As shown in **Figure 3**, the dominating processes responsible for the spectral shape are the photoelectric effect and Compton scattering. The influence of multiple scattering is negligible as variation of *prodfac* parameter to extreme settings does not show significant influence on the spectral shape. The effect of the geometrical positioning of the source was also studied. A mispositioning does not lead to changes in the spectral shape but significantly influences the total count rate, as would be expected because the incidence opening angle is varied through this procedure.

From these simulations it was concluded that the simulation is inherently stable towards parameter variation at a reasonable scale and the activity normalization is sensible. The observed discrepancies especially in the continuum close to the last peak could not be resolved by altering the physics parameters and must therefore be either of experimental origin or due to geometrical properties of the experiment not properly accounted for in the simulation.

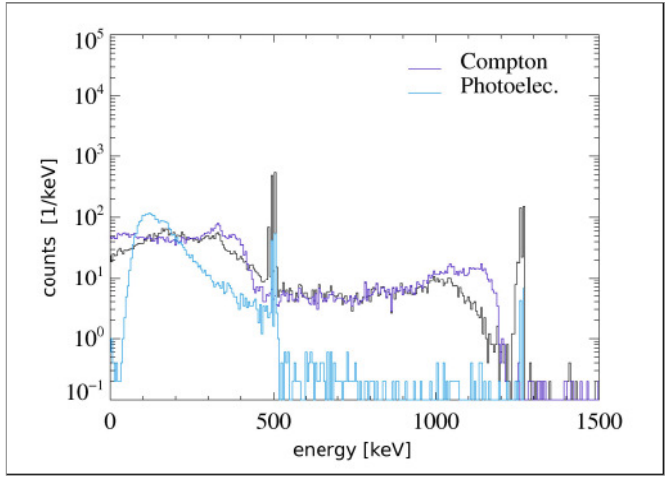

**Fig. 3** Processes contributing to secondary electron production in the detector bulk due to the incident gamma rays.

**2. Overall Representation of the Experiment by the Simulation**

As exemplarily shown for $^{22}$Na in **Figure 4** especially the continuum is not always correctly modeled by the simulation. Also the peak location and height is not always correct as can be seen by the oscillations in the deviation plot.

An analysis of the underlying database revealed that a number of gamma levels were wrongly included as shown in **Figure 5**, but this is not the case for the shown $^{22}$Na source. Instead geometrical effects such as dead layers or the influence of walls and ceiling may be of significance and will be tested as part of the ongoing work.

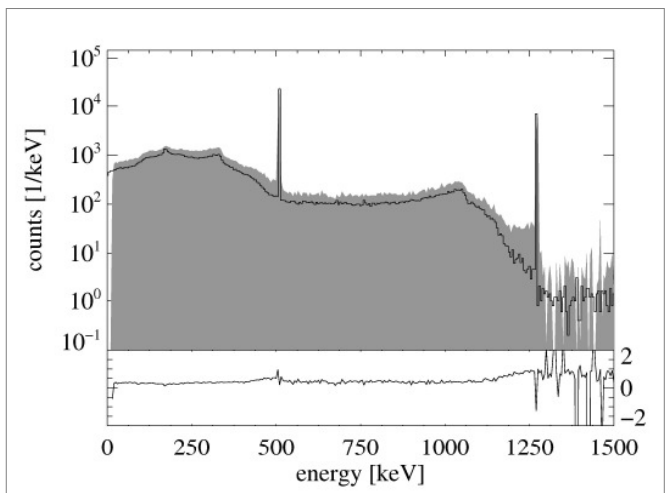

**Fig. 4** Experimental (filled) and simulated spectrum of $^{22}$Na source. The continuum is underestimated in the simulation and the peak to continuum ratios are not modeled correctly.

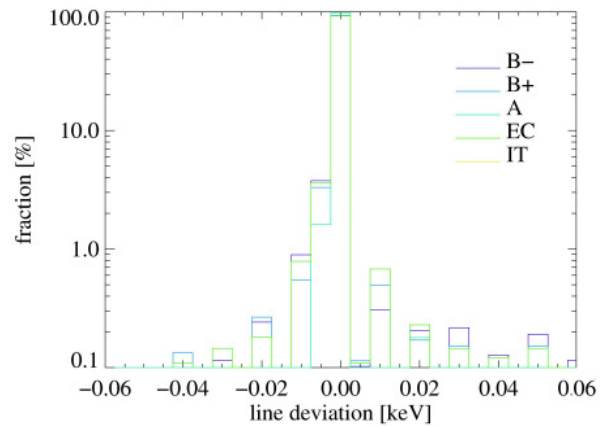

**Fig. 5** Difference of energy levels for all isotopes included in the Geant4 radioactive decay database with respect to the values given in the ENSDF.

### III. Conclusion

We have shown that the Geant4 radioactive decay module is capable of qualifiedly reproducing the prominent features of a simple gamma spectroscopy experiment. We have also shown that an accurate representation of the continuum and the peaks is not easy to achieve and requires iteration towards the optimal geometry. Since this is not possible in the planning phase of a new experiment, we conclude that for the Geant4 radioactive decay simulation to be a reasonably accurate tool for estimating the decay background in high sensitivity experiments further validation needs to be performed. We also see the need for a modern implementation of the underlying database which is quickly verifiable against established sources. As a side effect a modern implementation may also significantly increase the overall speed if modern programming techniques such as templating are used.